\documentclass[sigconf]{acmart}

\AtBeginDocument{%
  \providecommand\BibTeX{{%
    \normalfont B\kern-0.5em{\scshape i\kern-0.25em b}\kern-0.8em\TeX}}}

\usepackage[T1]{fontenc}
\usepackage[utf8]{inputenc}
\usepackage{float}
\usepackage{tcolorbox}
\usepackage{fontawesome5}
\usepackage{comment}
\usepackage{multirow}
\usepackage{amsthm}
\usepackage{graphicx}
\usepackage{longtable}
\usepackage{tabularx}
\usepackage{changepage}
\usepackage{url}
\PassOptionsToPackage{normalem}{ulem}
\usepackage{ulem}
\usepackage{array}
\usepackage{booktabs}
\usepackage{amsmath}
\usepackage{stackrel}
\usepackage{amsmath}
\usepackage{amsthm}
\usepackage{pgfplots}
\usepackage{pgf-pie}
\usepackage{balance}
\usepackage{hyperref}
\usepackage{fontawesome5}
\usepackage{subcaption}
\usepackage{pgfplots}
\usepackage{pgfplotstable}
\pgfplotsset{compat=1.7}
\usepackage{tikz}

\hypersetup{
  colorlinks   = True, 
  urlcolor     = blue, 
  linkcolor    = blue, 
  citecolor    = blue 
}

\copyrightyear{2023}
\acmYear{2023}
\setcopyright{acmcopyright}\acmConference[XXX 'xx]{XXX}{xx XX--XX,
XX}{XX, XX}
\acmBooktitle{XX), XX X--X, XXXX, XX, XX}
\acmPrice{15.00}
\acmDOI{XXXX}
\acmISBN{XXXX}

\begin{document}
\title[Agile Practices for  Quantum Software Development: Practitioners' Perspectives]{Agile Practices for  Quantum Software Development: Practitioners' Perspectives}


\author{Arif Ali Khan$^{1*}$, Muhammad Azeem Akbar$^{2}$, Aakash Ahmad$^{3}$ \\Mahdi Fahmideh$^{4}$, Mohammad Shameem$^{5}$, Valtteri Lahtinen$^{6}$, Muhammad Waseem$^{7}$, Tommi Mikkonen$^{7}$}
\affiliation{%
 \institution{$^{1}$M3S Empirical Software Engineering Research Unit, University of Oulu, Oulu, Finland}
\institution{$^{2}$Department of Software Engineering, LUT University, Lappeenranta, Finland} 
\institution{$^{3}$Lancaster University Leipzig, Germany}
 \institution{$^{4}$School of Business at University of Southern Queensland, Queensland, Australia}
 \institution{$^{5}$Department of CSE, Koneru Lakhmiyah Educational
Foundation, Vaddeswaram, Andhra Pradesh, India}
\institution{$^{6}$Quanscient Oy, Tampere, Finland}
\institution{$^{7}$University of Jyvaskyla, Finland}
 \institution{arif.khan@oulu.fi, azeem.akbar@ymail.com, a.ahmad13@lancaster.ac.uk, mahdi.fahmideh@usq.edu.au\\ shameem.ism@gmail.com, valtteri.lahtinen@quanscient.com, wasimsse@gmail.com, tommi.j.mikkonen@jyu.fi} \country{}}
\renewcommand{\shortauthors}{Khan et al.}

\begin{abstract}
Quantum software systems are emerging software engineering (SE) genre that exploit principles of quantum bits (Qubit) and quantum gates (Qgates) to solve complex computing problems that today’s classic computers can not effectively do in a reasonable time. According to its proponents, agile software development practices have the potential to address many of the problems endemic to the development of quantum software. However, there is a dearth of evidence confirming if agile practices suit and can be adopted by software teams as they are in the context of quantum software development. To address this lack, we conducted an empirical study to investigate the needs and challenges of using agile practices to develop quantum software. While our semi-structured interviews with 26 practitioners across 10 countries highlighted the applicability of agile practices in this domain, the interview findings also revealed new challenges impeding the effective incorporation of these practices. Our research findings provide a springboard for further contextualization and seamless integration of agile practices with developing the next generation of quantum software.
\end{abstract}

\ccsdesc[500]{Software and its engineering~Designing software}
\ccsdesc[500]{General and reference~Empirical studies}
\keywords{Agile Practices, Quantum Software Engineering, Quantum Computing}

\maketitle       
\section{Introduction}
\label{sec:introduction}
The field of Quantum computing (QC) has received rapidly growing industrial and policy interest to an even greater extent. It is expected to bring revolution in various industrial areas \cite{zhao2020quantum}. It is evidenced by the fact that technology giants such as IBM \cite{qiskt2021}, Google \cite{cirq2022}, and Microsoft \cite{azure2021} have heavily invested in implementing QC as a service offering to tackle a new class of complex computational problems. Developing QC applications, on the other hand, is a challenging endeavour. The literature suggests that QC development is, after all, essentially a type of system development endeavor \cite{zhao2020quantum}. Given this analogy, adopting a systematic engineering lifecycle perspective for managing the complexity of QC development is acclaimed \cite{zhao2020quantum}. This takes precedence over an ad-hoc use of implementation techniques and technologies or relying on the development skills of the individuals that may likely to deliver QC, which may be erroneous and costly to maintain \cite{khan2022software}\cite{zhao2020quantum}. Hence, to achieve an optimum effect, there must be advanced SE Methods to enable QC to live up to its promising potential. 

The key enabler for developing QC technologies is quantum software \cite{ali2022software}. Quantum software, amongst other requirements such as full stack support ranging from novel techniques and tools, needs processes and methods that explicitly focus on developing software system based on quantum mechanics \cite{ali2022software}. However, the present-day quantum software engineering (QSE) processes are far from being mature, and are most often based on hybrid concepts (consisting of quantum-classical tools and practices). The quantum-classical development must be orchestrated; therefore, we previously presented the vision of embarrassing  iterative and agile practices for developing a quantum software \cite{khan2022embracing}. The QSE processes can benefit from well-developed iterative agile practices for team collaboration, short development iterations, and continuous delivery \cite{beck2001manifesto}\cite{fitzgerald2006customising}. Presently, adopting a more “agile” approach to develop a quantum software system is  reliable option rather than waiting for domain specific QSE processes and methods \cite{ali2022software}. However, no empirical evidence yet exists that ties agile practices and quantum software development activities. This study aims at exploring the significance of  agile practices in the QSE domain. 

 We conducted semi-structured interviews with a total of 26 software practitioners (having mutually inclusive knowledge of agile software engineering and quantum software development) across 10 countries with diverse roles such as quantum algorithm developer, software developer, agile coach working in varying domains ranging from cyber security to automotive and healthcare, allowing us to gain different perspectives on our research aim. To analyse interview results, we adopted the Grounded Theory (GT) method \cite{glaser2017discovery} in a bottom-up approach to derive a mapping of code, concepts, and categories from raw (interview) data. 
 
 The results indicate that practitioners view  agile practices as the best fit to support quantum software development activities. However, small portion of the interview participants find themselves underprepared due to lack of a basic knowledge of quantum mechanics, insufficient expertise and non-availability of tools, that hinders the progress of agile-driven quantum software development.  
 
 We outline the primary contributions of this research as (i)  empirically-derived understanding on the adoption of agile in quantum software development, (ii) discussing four major categories of challenges \textit{(knowledge and awareness, sustainable scaling, quantum-aware tools and technologies} and \textit{standards and specifications)} that must be addressed for adopting agile practices in quantum software development domain. The study results in principle aim to enlighten software researchers and practitioners by pinpointing the understanding and challenges of using agile practices for developing a quantum software.
 
The rest of the paper is organised as follows. The background of the study is described in section \ref{sec:background}. The research methodology process is discussed in section \ref{sec:methodology}. Section \ref{sec:Analysing Interview Data} detail the data analysis process and   results are presented in section \ref{sec:Results}. The discussions, study implications and threats to the validity of study findings are reported in section \ref{sec:discussion}. Section \ref{sec:relatedWork} reviews the related work and section \ref{sec:conclusions} draws the conclusions and outlined the future directions.

\section{Background} \label{sec:background}
Firstly, we contextualise the development of software-intensive systems that can be executed or simulated on QC platforms. The concepts and terminologies introduced in this section, illustrated in Figure \ref{fig:QSE} are used throughout the paper to elaborate technical aspects of SE for QC. 

\subsection{Quantum Software Engineering}
\begin{figure}[t]
 \centering
 \includegraphics[width=0.3\textwidth]{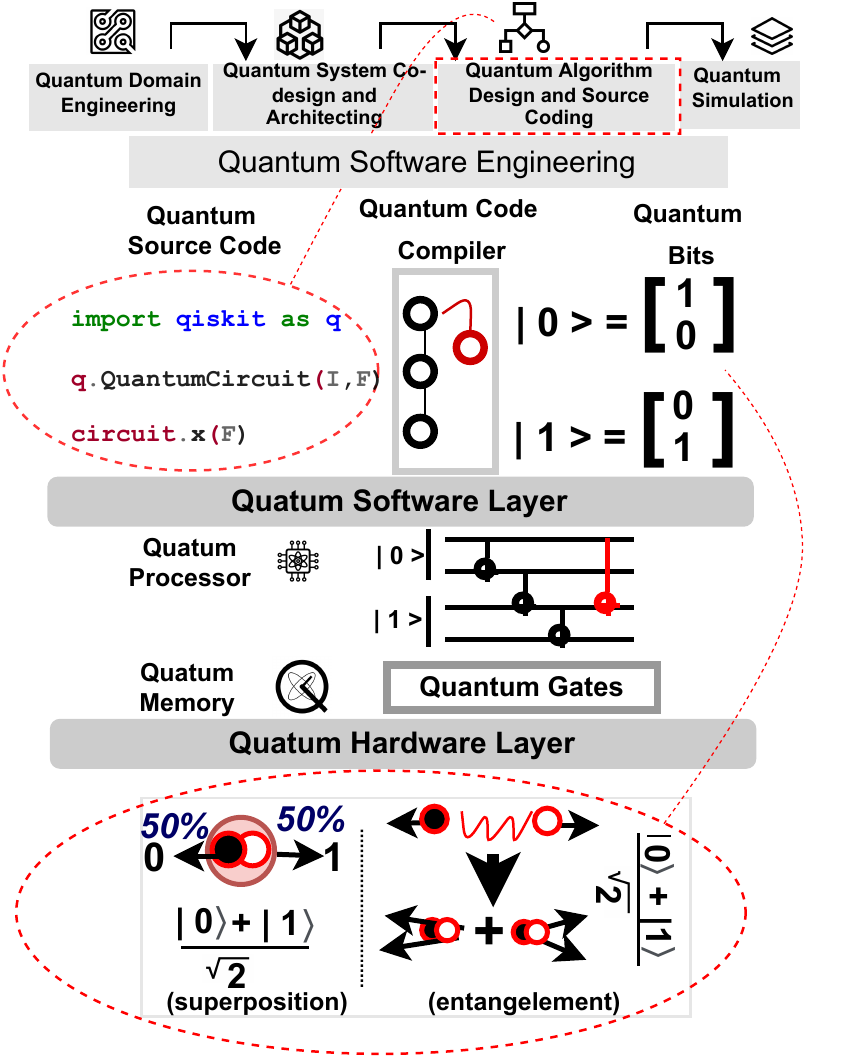}
\caption{Overview of quantum computing}
\label{fig:QSE}
\end{figure}

Quantum computers harness the phenomena of quantum mechanics, e.g., quantum superposition and entanglement to process, store, and transmit quantum information set represented with Qubits that manipulate Qgates, as in Figure \ref{fig:QSE}. In contrast to classic binary digits represented as (0,1), Qubit as the most fundamental unit of quantum information set attains a state that is a superposition of 0 and 1, and represented as |0⟩ and |1⟩\cite{zhao2020quantum}\cite{rieffel2011quantum}, where:
\begin{equation}\label{EQ-1}
 |0\rangle  =  \left[ \begin{array}{c} 1 \\ 0 \end{array} \right] ~~~~~ \hspace{1cm}  \hspace{1cm} ~~~~~  |1\rangle  =  \left[ \begin{array}{c} 0 \\ 1 \end{array} \right]   
\end{equation}

From computation (i.e. QC) and software system development (i.e.QSE) perspective, quantum computer disrupt the idea of traditional digital systems and represent a paradigm shift (transitioning from bits to Qubits) with a vision to next generation computing \cite{zhao2020quantum}. In a QC context, operationalising Qubits that manipulate quantum circuits (Qgates as a primitive unit of quantum hardware), software programmers require quantum age algorithms and programming languages such as Microsoft Q\# \cite{azure2021}, IBM Qiskit \cite{qiskt2021} and Cirq by Google \cite{cirq2022}. Quantum programming language and their underlying algorithms are well suited to focus on implementation details that produce executable specifications on quantum computer; however, they lack an engineering view for developing quantum software including the activities such as quantum domain engineering, quantum system co-design, quantum algorithm design and source coding and quantum information simulation (see Figure \ref{fig:QSE}) \cite{khan2022embracing}\cite{ahmad2022towards}. QSE as the most recent genre of SE aims to exploit the status-quo (ISO/IEC/IEEE 12207:2017 standard) \cite{ISO/IEC/IEEE2017} by leveraging the engineering processes, reference architectures, patterns, tools and framework to develop quantum-intensive software. There is no well established process or even a notion of how QSE process might look, however; some recent research efforts are seen as attempts to find a common denominator of engineering activities, highlighted in Figure \ref{fig:QSE}, to establish foundations for process-centric QSE e.g. \cite{weder2021quantum}\cite{khan2022embracing}\cite{ahmad2022towards}. Building on the notion of a QSE intensive process, new software development processes and methods or customizing the existing conventional methods (hybrid quantum-classical) are required \cite{weder2021quantum}\cite{hernandez2020quantum}\cite{khan2022software}.

\subsection{Agile for Quantum Software Development}
In line with the legacy of experiences in conventional programming, which also started from hardware-focused, hard-wired techniques in 1950s and then evolved into today's agile system development practices, the QSE should eventually follow the same agile development tradition \cite{moguel2020roadmap}\cite{piattini2021toward}. However, tools and methods that are used to achieve agility in conventional software engineering need to be examined in line with the characteristics of quantum software development \cite{khan2022embracing}. When coding quantum programs, software developers face new challenges due to switching to an entirely different programming mindset with counterintuitive quantum principles \cite{khan2022embracing}. For example,  executing the quantum instructions in the state of qubits and measuring the qubit values \cite{de2022software}.

Testing quantum programs requires at least the development of solutions to tackle the following challenges of (1) defining test oracles as the state of a quantum program can be in superposition and thus difficult to get precise state; (2) running efficient quantum test data generation since quantum variables might be exponentially higher than classical variables; and (3) dealing with false positives/negatives due to vulnerabilities to hardware glitch \cite{wang2022generating}\cite{wang2022mutation}.
Finally, faults found in quantum programs with testing must be located, isolated, and patched; thus, debugging is needed. Furthermore, developing effective debugging solution is impeded by the following challenges of: (1) examining values of quantum variables in superposition; (2) interpreting multi-dimensional quantum states, despite the availability of simulation techniques, and (3) lacking guidance on where and what to check when debugging quantum programs \cite{huang2019statistical}. 

In conventional software engineering, the above mentioned challenges are alleviated via adopting the agile and iterative principles. In line with this, there has been a long-standing acknowledgment that agile practices have shown their efficacy in practice \cite{dybaa2008empirical}. Proponents of the agile practices claim, in contrast to the traditional methodologies (e.g. Waterfall model) \cite{dybaa2008empirical}, agile solves many of the problems endemic to the field by proposing principles and practices such as active user involvement, short iterations, small and frequent release, and refactoring  \cite{beck2001manifesto}. This, in turn, supports their state of flow, which is an important ingredient in modern-day software development \cite{mikkonen2016flow}. 
In present scenario, QC is still in its infancy, and the situation of developing a quantum software could considerably be eased by adopting agile practices as suggested by Piattini et al. \cite{piattini2021toward}: \textit{"we should adopt a more “agile” approach when
proposing and developing software engineering
quantum techniques, that is, do not wait until
quantum programming languages are “stable” or
“refined” in order to adapt existing techniques or create new ones, but develop them in parallel with the evolution of the quantum languages, starting from now"}. Agile software development help find bugs and other problems as early as possible because at that point the developers can fix them with straightforward actions. Moreover, today quantum software development activities are mostly hybrid (classical and quantum) \cite{weder2021quantum}, therefore conventional agile practices provide a set of best practices for developing a quantum software through collaborative efforts of self-organising and cross-functional teams. Gonzalez and Paradela \cite{hernandez2020quantum} complemented it by explicitly mentioning that: \textit{"Quantum software development projects currently have numerous features that fit neatly into the agile paradigm, such as adding features evolutionarily or using trialand-error algorithms"}.  

However, to the best of our knowledge, no empirical study has yet been conducted to explore the perception of practitioners for using agile practices to develop a quantum software system. This study aims to fill the given research gap by conducting interviews with practitioners to address the core research questions (RQs) discussed in the section \ref{Sec:RQs}.

\section{Methodology} \label{sec:methodology}
We now detail the research methodology, as illustrated in Figure \ref{fig:methodology}, which comprises the following steps.
\subsection{Context (Process-centred quantum software)
}
\begin{figure*}[t]
 \centering
\includegraphics[width=0.73\textwidth]{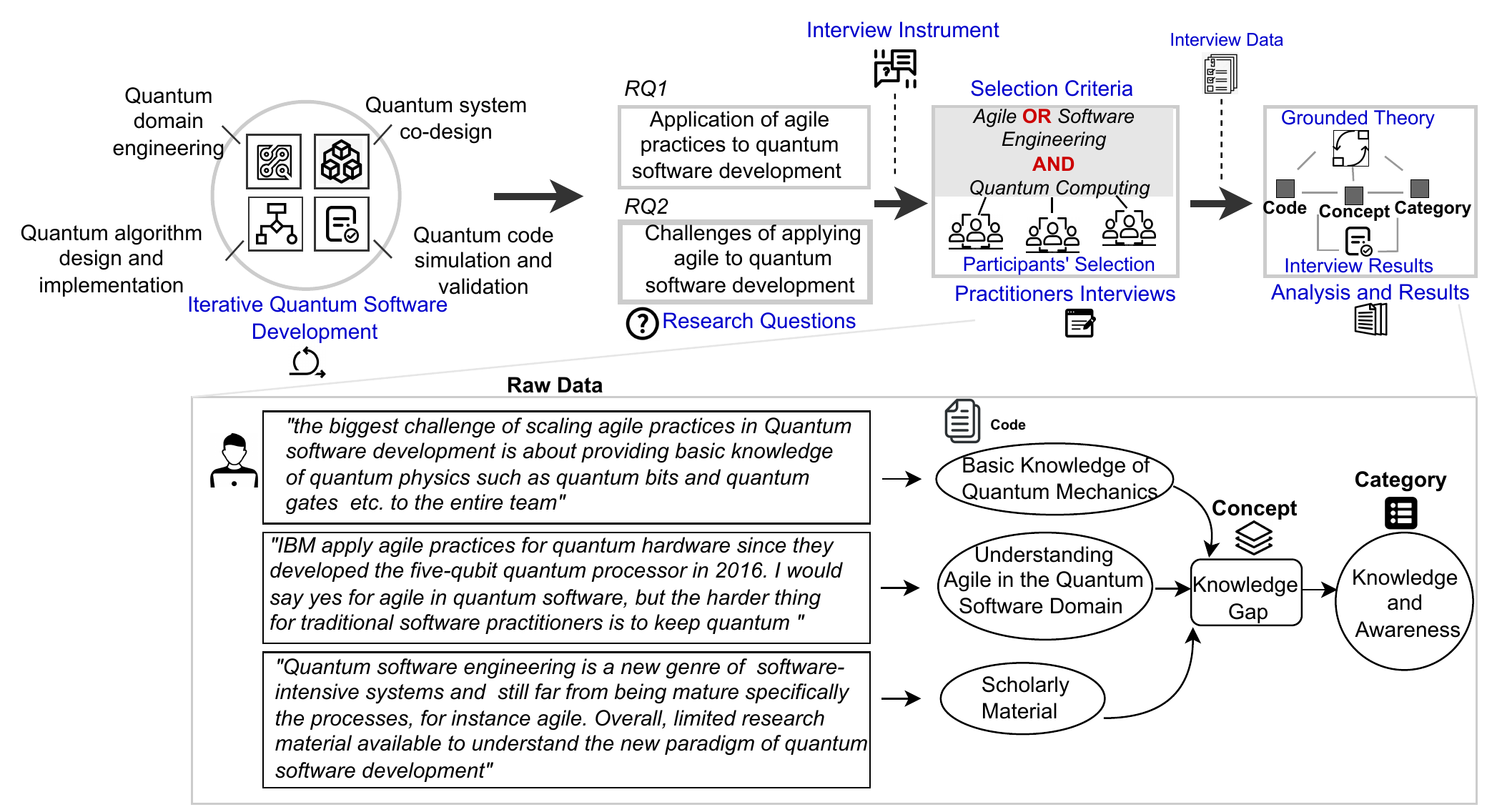}
\caption{Overview of the research methodology}
\label{fig:methodology}
\end{figure*}

To understand a structured approach for developing quantum software, we collaborated with industrial partner (Quanscient Oy)\footnote{\url{https://quanscient.com/}} and studied the incremental quantum software development process \cite{khan2022embracing}  - including four iterative activities (i) quantum domain engineering (ii) quantum system co-design  (iii) Quantum algorithm
design and implementation, and (iv) quantum code simulation and validation (see Figure \ref{fig:methodology}). For illustrative purpose, Figure \ref{fig:methodology} demonstrates what iterative activities needs to be done and each activity details how it is to be done.  

However, this study aims to extend our previous work \cite{khan2022embracing} by using agile practices to support the quantum software development process activities. The findings of this study will provide in-depth understanding of agile-driven quantum software development. Therefore, we structured the following research questions and developed the interview instrument to achieve the study aims.

\subsection{Research Questions} \label{Sec:RQs}
We formulated following two research questions (RQs) that help us to present the results of the study.

\textbf{RQ1}: Do agile practices best suited for developing quantum software?

\textbf{Rationale}: To understand the practitioners’ perceptions on the significance of agile practices in quantum software development. It has yet to explore if agile practices are applicable, or may need augmentations to be applicable for developing quantum software systems \cite{khan2022embracing}.

\textbf{RQ2}: What are the challenges of adopting agile practices for developing a quantum software system?

\textbf{Rationale}: To identify the challenges that impede adopting agile practices for developing quantum software. Identification and classification of the challenges can help to explore potential areas which need significant academic and industrial attention for defining guidelines and/or solutions that address or mitigate these challenges.

\subsection{Interview} \label{sec:InterviewMetho}
\underline{\textbf{Interview instrument}}: Following the guidelines by Robinson \cite{robinson2014sampling}, we developed semi-structured questions which covered three categories, including (i) demography and professional details, (ii) suitability of agile practices for developing quantum software (RQ1), recorded via a 5-point Likert scale (Strongly agree to Strongly disagree) detailed later and (iii) challenges of agile-driven quantum software development as open-ended question for spontaneous responses (RQ2) (analysed with GT in Figure \ref{fig:methodology}). The first three authors conducted regular meetings and developed the interview questions; finally, a Zoom meeting  was called and invited all other authors to conclude the interview instrument\footnote{\url{https://tinyurl.com/5y2fd56m}}.

\underline{\textbf{Recruiting participants}}: We strictly sought practitioners with a mutually inclusive knowledge of ‘agile SE’ and ‘QC’ ensured via contacting the potential participants and informal discussion on their willingness and required basic knowledge of the subject. We used relevant platforms such as GitHub, professional social media networks (i.e., LinkedIn, ResearchGate, WeChat, Facebook, mailing groups), industrial network and personalised emails to contact the targeted population. A total of 73 practitioners were contacted and eventually 26 participated in the interview as shown in Figure \ref{fig:demography}(a).

\underline{\textbf{Pilot interviews}}: Three pilot interviews were conducted to seek early feedback on the types and formulation of questions, the required time to complete the interview, and the assessment for collecting interview data. Note that pilot interviews data are not used for the final data analysis. It is only used to refine the understandability, readability and structure of the interview questions. 

Based on the refinement from pilot interviews, we  shared the final interview questions with 26 agreed participants a week before the session and provided the author's contacts if they had any questions regarding the interview material. The interview sessions were conducted online by the second author using Zoom, VooV meeting and Microsoft Teams platforms.

\section{Analysing Interview Data} \label{sec:Analysing Interview Data}
Based on the semi-structured interview questions (Section \ref{sec:InterviewMetho}), we analysed the data in three steps. First step analysed the demographic details of the practitioners such as their country, years of experience, professional domain etc., as presented in Figure \ref{fig:demography}. Demography details complement the analysis of interview data corresponding to RQ1 and RQ2. For example to analyse if factors like years of experience, domain of experience (cyber security or automotive engineering etc.) and/or professional roles (agile coach or quantum algorithm developer etc.) impacts practitioners perspective or knowledge on the application of agile practices to quantum software development. 

Second step involved recording the interview data for RQ1, using a 5-point Likert scale \textit{(Strongly Agree, Agree, Neutral, Disagree, Strongly Disagree)} followed by an open-ended question to rationalise or describe their choice, as shown in Figure \ref{fig:RQ1}.

The last step focused on RQ2 and we adopted the Grounded Theory (GT) method \cite{glaser2017discovery} to derive a mapping of code, concepts, and categories as in Figure \ref{fig:methodology}.  GT is a systematic method to generate a theory or conceptual view out of raw data and to enable researchers to have amenable interpretation and free comprehension of data in different ways \cite{glaser2017discovery}. We adopted the GT approach for various reasons, e.g., agile practices are more people-oriented, and GT helps to examine teams' behaviors and social interactions explicitly \cite{hoda2010organizing}. Similarly, Hoda et al. \cite{hoda2010organizing} mentioned that GT is the best fit for novel research fields which have not been previously well explored, and the research on agile in quantum software development is scarce. GT has received growing attention in software engineering research and more narrow in the agile domain e.g. \cite{madampe2021faceted}\cite{masood2020real}\cite{hoda2017becoming}.

The GT open coding \cite{allan2003critique} and constant comparison approaches \cite{glaser2017discovery} were followed to analyze the raw data and identify the core categories of agile-driven quantum software development challenges. The data coding and mapping are preliminary conducted by the first three authors following the GT guidelines \cite{glaser2017discovery}. However, the rest of the authors were invited to participate and provide feedback in the final data coding and concept development process.
The following raw data example is presented to explain the analysis process stage to coding, concept development, and categorization (Figure \ref{fig:methodology}).

\underline {\textbf{Raw data}}: 
\faComment{} \textit{"the biggest challenge of scaling agile practices in Quantum software development is about providing basic knowledge of quantum physics such as quantum bits and quantum gates  etc. to the entire team"} [P17], Quantum sofwtare engineer. 

\underline {\textbf{Code}}: \textit{Basic knowledge of quantum mechanics}

The codes evolved from each interview response were constantly compared with the codes of the same interview and the entire data set, including the interviews of other participants. The mentioned code \textit{"basic knowledge of quantum mechanics”} (P17, Quantum software engineer) was found similar to two other codes, namely \textit{"understanding agile in the quantum software domain”} (P10, Scrum master) and \textit{"scholarly material”} (P14, Agile trainer). The similar codes were grouped and defined the common concept, which is the higher level of abstraction.

\underline {\textbf{Concept}}: \textit{Knowledge gap}
 
The other relevant concepts that emerged during the constant comparison of codes are the  \textit{"team dynamics"} and \textit{ "quantum software development education"}. The final constant comparison was conducted for a similar group of concepts to create the final, and high level of abstraction called categories.  

\underline {\textbf{Category}}: \textit{Knowledge and awareness}

Knowledge and awareness is a high-level concept that encapsulates the insights and perceptions of participants, who were more concerned about the guidance, support and awareness of quantum computing concepts and their integration with agile principles.

In this study, the level of abstraction (coding, concept creating, category development) was based on real-world data; therefore, the subsequent discussion is grounded in the context of the collected data. Note that the other categories of challenges were derived following the same open coding and constant comparison methods. However, we provided a sample analysis process only for the knowledge and awareness category but omitted it for all the other categories because of space restrictions. The emerged categories represent the challenging of agile-driven quantum software development. We are not claiming the universal generalizability of the study results grounded on the collected data; however, it perfectly described the investigated context \cite{adolph2008methodological}.

The categories constituting \textit{knowledge and awareness, sustainable scaling, quantum-aware tools and technologies} and \textit{standards and specifications} along with the important quotes from practitioners are delineated in the following sections. We further considered the developed concepts to support the findings; however, the space restrictions limited us from describing them in detail.

\section{Results}\label{sec:Results}
\begin{figure*}[t]
 \centering
 \includegraphics[width=0.7\textwidth]{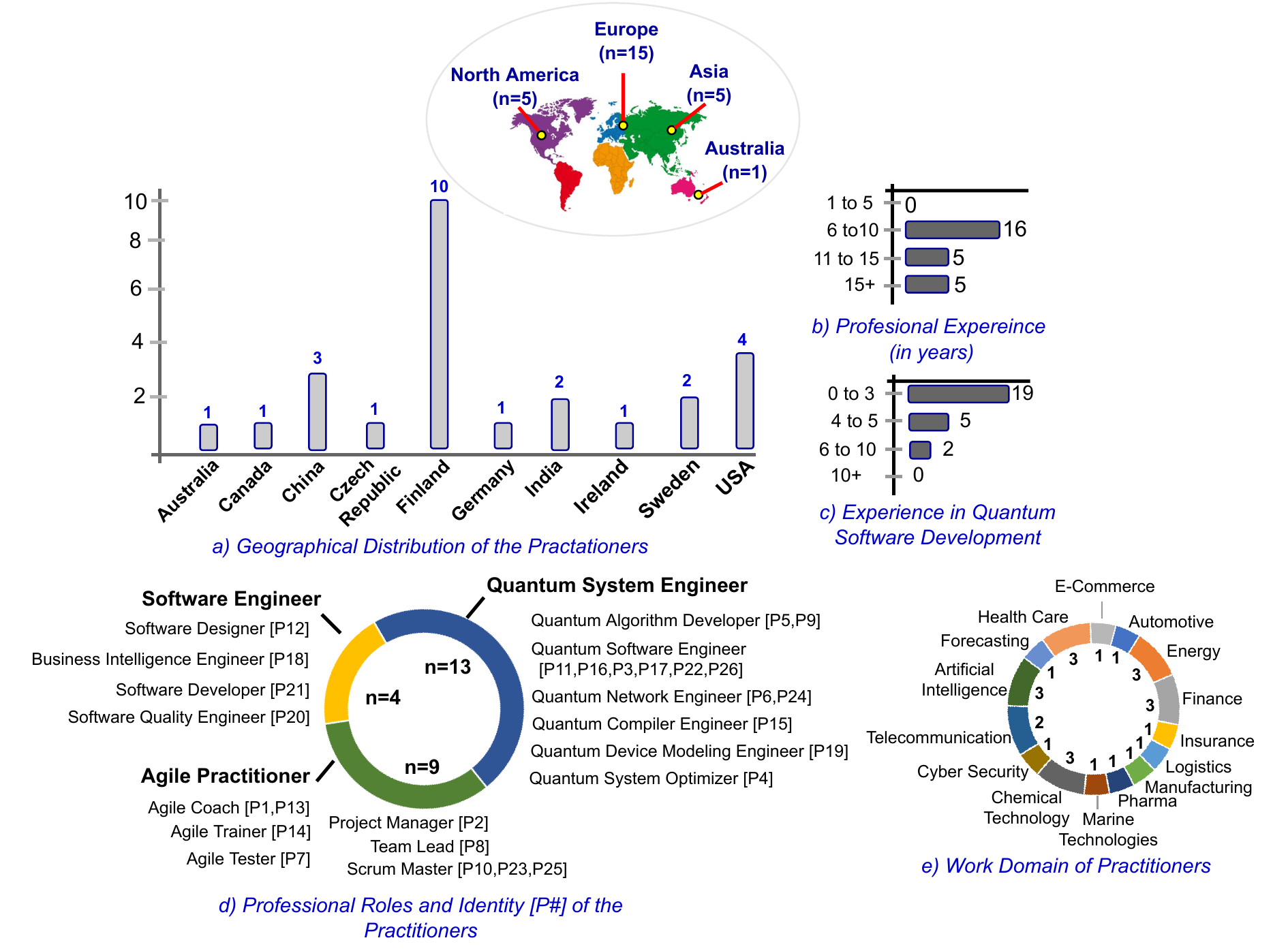}
\caption{Overview of interview participants demographics}
 \label{fig:demography}
\end{figure*}

This section presents the study results, addressing the two RQs outlined in Section \ref{Sec:RQs}. 
Moreover, the demographic analysis was conducted to examine the dimensions and dynamics of the targeted population and contextualize the responses that complement agile-driven quantum software development for a specific group of practitioners (see Figure \ref{fig:demography}). We noticed that 26 respondents from 10 countries across 4 continents with 16 roles and 15 different work domains participated in this study (see Figure \ref{fig:demography}(a,d,e)). Note that each interview participant is tagged with a unique id [P\#], as shown in Figure \ref{fig:demography}(d). The experience of interview participants in the quantum software development domain mostly ranges from 0 to 3, which is (n=19) of the total responses (see Figure \ref{fig:demography}(c)). Of all the responses, the majority (n=16) have 6-10 years of professional working experience as practitioners (see Figure \ref{fig:demography}(b)). The results illustrate that most (n=13) of the interview participants work as quantum system engineers, which is further classified across 6 different sub-roles (see Figure \ref{fig:demography}(d)). Therefore, the given demographic findings reveal that the interview participants are diverse with respect to geographical locations, experiences, roles, and industrial domains. It gives us the confidence to generalize the study findings to some extent. The detailed results to address both RQ1 and RQ2 are presented in the following sections.

\subsection{Agile for Quantum Software Development (RQ1)}\label{Sec:ResultsRQ1}
\begin{figure}[t]
 \centering
  \includegraphics[width=0.31\textwidth]{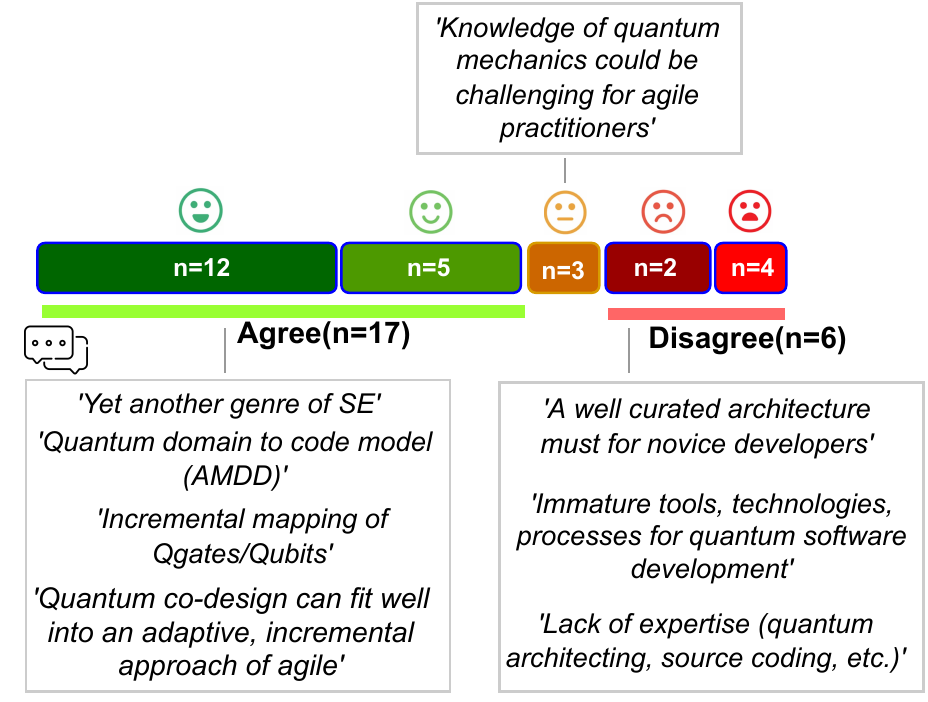}
\caption{Agile practices for quantum software development}
\label{fig:RQ1}
\end{figure}

To understand practitioners’ perspectives on the applicability of agile practices to develop quantum software, we sought their feedback using a five-point Likert scale as shown in Figure \ref{fig:RQ1}. The selection of a five-point pre-defined response was complemented by an open-ended optional question that allowed practitioners to express the description of their choice spontaneously. Figure \ref{fig:RQ1} shows that a total of 12 practitioners agreed strongly, and 5 agreed (i.e., n=17) that agile practices can be used to develop quantum software. The open-ended descriptions suggest that practitioners perceive QSE as another genre of software engineering, domain modeling, incremental mapping of Qgates and Qubits, and quantum system co-design that can fit well in agile software development. For example, a participant suggested that s/he strongly agrees on the application of agile practices to quantum software development as quantum system co-design, i.e., translating the quantum software requirements into quantum design and source code can benefit from the incremental approach of agile. 
On the other hand, three participants  remained neutral, rationalizing their view on knowledge of quantum mechanics (operationalizing Qubits) can be challenging for traditional software engineers (see Figure \ref{fig:RQ1}). Furthermore, 2 participants strongly disagreed, while 4 disagreed (i.e., n=6) on application of agile for quantum software development. They justified the counterproductivity of applying agile practices in quantum software development by reasons such as immature tools and technologies, lack of professional expertise, and a well-curated architecture.  
We conclude that while answering RQ1, approximately 2/3 majority recommended the application of agile practices to develop quantum software, an incremental mapping of quantum domain concepts can help to operationalize quantum software. Approximately 1/4 of practitioners cited as non-availability of tools, and professional expertise can be a reason for their disagreement.

\subsection{Challenges of Agile-driven Quantum Software Development (RQ2)} \label{sec:ResultsRQ2}
The identified five key categories \textit{(knowledge and awareness, sustainable scaling, quantum-aware tools and technologies} and \textit{standards and specifications}) present the challenging aspects that need to consider for scaling agile practices in the quantum software development domain (see section \ref{sec:Analysing Interview Data}). The following sections describe each category in detail with the quote support selected from the interview data.
\subsubsection{Knowledge and awareness}
The awareness and knowledge of agile practices in the quantum software development domain have emerged as an important category, consisting of three concepts: knowledge gap, team dynamics, and quantum software development education.
 \begin{itemize}
     \item \textbf{Knowledge gap}
 \end{itemize}

Agile practitioners most likely experience some particular areas of quantum software development more challenging because of the knowledge gap (classical->quantum), such as quantum domain engineering, architecting quantum software, quantum source code interpretation, and quantum algorithm complexity \cite{ahmad2022towards}\cite{li2021understanding}. For instance, the conventional agile practitioner might not know how to design a quantum algorithm attuned to operationalize Qubits rather than classical bits \cite{ahmad2022towards}. 

The feedback of interview participants is aligned with it and perceived that knowledge of quantum computing helps practitioners in the better adoption of agile practices as endorsed by one of the interviewee: 

\faComment{} \textit{"But as this wave of innovation continues, a necessary element lags behind: a quantum-skilled workforce. To develop  quantum software workforce, the agile iterations give quick feedback and chance of speed learning and experiences"} [P10], Scrum master.

Generally, agility refers to delivering customer value faster, incremental, and with less headaches \cite{beck2001manifesto}\cite{dybaa2008empirical}. Despite this, managing agile activities in the quantum software domain  considered with worries \cite{khan2022embracing}. Participants recommended the need for the alignment among skills, quantum awareness, resources, and technologies for developing agile-based quantum software, as mentioned by an interview respondent. 

\faComment{} \textit{“Agile teams can get bogged down by failure in the novel quantum paradigm. However, starting with the end goal is imperative when building an agile team to work on a quantum software engineering project. How will you reach your goal, and what skills, technology, and hours will you need? Only then you can build the team”} [P13], Agile coach.

\begin{itemize}
     \item \textbf{Team dynamics}
 \end{itemize}
Agile practices have inherent focus on team structure (e.g., common goals, roles, responsibilities, collaboration), and organized team dynamics is a key to high functional agile productivity \cite{Intelliware2014} \cite{beck2001manifesto}\cite{dybaa2008empirical}. A happy agile team will essentially be more productive; conversely, the team can be extremely ineffective \cite{Intelliware2014}. Agile team dynamics is challenging to manage in the quantum software domain because of the different quantum rules set. The agile team’s mindset will likely change due to quantum physics characteristics such as superposition, entanglement, and quantum interference, which significantly influence the way software development team works \cite{khan2022embracing}. This rationale was captured quite well by a respondent:

\faComment{}\textit{“Because of flaky characteristics of QC, agile teams face significant challenges due to switching to an entirely different programming paradigm. The bits and qubits interpretations and manipulation bring high-level confusion for agile team forming, storming, norming, and performing”} [P16], Quantum software engineer.

\begin{itemize}
     \item \textbf{Quantum software development education}
 \end{itemize}
Quantum software development as a computer science subject is still at an early stage \cite{peterssen2020quantum}- in contrast to QC education, which is already taught in various universities \cite{carberry2021building}. The increasing boom of QC technologies calls for educational approaches that structure core ideas and terms for quantum software development as a subject \cite{peterssen2020quantum}. The respective applications of core ideas constitute the foundation for defining quantum software development curricula. The above literature findings were supported by the respondents more narrow in an agile-quantum context, for example:

\faComment{} \textit{“Academic treatment require to offer agile-enabled quantum software development as a subject area, at least at the university level. It will give learning opportunities to realize the capabilities of agile practices for developing quantum software and allow students to exercise their cognitive muscles as well as joined-up thinking. I understand that incorporating agile-based quantum software engineering as a subject in degree programs will help in preparing a skilled workforce to fulfill the future needs of the quantum software industry”} [P20], Software quality engineer.

\subsubsection{Sustainable scaling} \label{sec:sustainable scaling}
Emerging quantum software engineering is a paradigm shift that demands enhancing the conventional processes and practices to support requirements of quantum reality \cite{zhao2020quantum} \cite{khan2022software}. Piattini et al. \cite{piattini2021toward} recommended scaling the conventional agile practices to develop quantum software; however, it is essential to set sustainable strategies to avoid potential risks and scaling pitfalls \cite{sophia2022sustainability}. For instance, one of the interview participants mentioned that:  

\faComment{} \textit{“Organizational and team structure is a huge part of the hybrid agile-quantum process, and top management should be prepared for it. Remember, be fast, and sustainable in integrating agile with quantum software lifecycle by identifying the process aspects that need improvement else immediately find alternatives”} [P2], Project manager.

The sustainable scaling category is based on the two concepts: ethically aligned quantum software and agile-quantum ecosystem. 

\begin{itemize}
     \item \textbf{Ethically aligned quantum software}
 \end{itemize}
 Quantum software engineering is a new genre of software development, and the boom of QC from research to the business raises various ethical concerns \cite{saurabh2022ethical}. The QC transition threatens the existing ethical protections, e.g., security, and transparency \cite{sophia2022sustainability}. For instance, one of the interview participants justify it more in agile-driven quantum software scenario: 
 
\faComment{} \textit{“ I have never come across a framework used to determine the ethical concerns raised in agile-quantum software development. I argue that regulatory bodies shall step forward to consider the ethical risks of quantum theory applications and invest in a coordinated approach to govern the potential risks effectively"} [P18], Business intelligence engineer.

The interviewee’s statement is supported by Buchholz and Ammanath \cite{ammanath2022ethics} broadly in the quantum computing domain by mentioning that the ethical concerns raised by QC are still on the horizon, and not possible to tackle them immediately. However, it’s time that industrial and government stakeholders realize the trigger events and start thinking to develop the strategies \cite{ammanath2022ethics}. The existing classical frameworks could be the best fit to start  building ethical guidelines and models in the QC domain.

\begin{itemize}
     \item \textbf{Agile-quantum ecosystem}
 \end{itemize}
 QC ecosystem is growing beyond advancements in hardware, and people have started focusing on producing quantum software and business partnerships \cite{veronica2022ecosystem}. Reaping the business interests of quantum software enables teams to connect and collaborate from anywhere. However,   developing sustainable collaboration between teams and businesses is challenging as collaboration sits at the intersection of  computer science, quantum physics and applied mathematics \cite{aggie2022skills}. The collaboration could be complemented by presenting an ecosystem as highlighted by an interview participant more specific in agile based quantum software development:
 
\faComment{} \textit{“ Agile activities more focus on collaboration and communication across teams, customers and other stakeholders. On the other hand, quantum software development is a paradigm shift and the agile teams must need an ecosystem that help them to seamlessly collaborator with customers and stakeholders from other domains (e.g. quantum physics)” }[P11], Agile coach.



\subsubsection{Quantum-aware tools and technologies}
A wide range of tools are available to support quantum software development activities. For example, Qiskit \cite{qiskt2021} is a known open-source tool available as a service on IBM cloud for developing quantum software at module, circuit and algorithm levels. Similarly, Microsoft Azure Q\# and the Quantum Development Kit offers tools as a service for quantum software development i.e., designing the quantum algorithm, optimizing the solutions, and applying the optimized solutions across the Azure platform to see the real-world impacts \cite{azure2021}. Moreover, Khan et al. \cite{khan2022software} thoroughly discussed the existing toolchain support for architecting a quantum software system. However, the interview participants highlighted its lack for customizing agile practices to develop quantum software. 

 \faComment{}\textit{“Presently, it’s exceedingly difficult to develop quantum software that offers commercial-level benefits which lift economies. We need to present novel tools and techniques that support hybrid quantum-classical development, for instance, using the traditional (classical) agile practices to support quantum software development activities”} [P6], Quantum network engineer.

The quantum-aware tools and technologies category emerged from two underlying concepts: classic-quantum tailoring and continuous SE infrastructure. 

\begin{itemize}
     \item \textbf{Classic-quantum tailoring}
 \end{itemize}

It is commonly acknowledged that one-size-fits-all application of methods and processes is fallacious \cite{fitzgerald2006customising}. In line with it, classical techniques must be tailored to achieve the optimum effect in the quantum software domain \cite{weder2021quantum}\cite{zhao2020quantum}. The tailoring could be actualized using domain-specific tools and technologies \cite{fitzgerald2006customising}. The need for such tools and technologies continue as a persistent theme in quantum software development \cite{khan2022software}\cite{weder2021quantum}. 

In this study, classical-quantum tailoring is particularly related to customizing the traditional agile practices for quantum software. However, the interview participants highlighted the lack of tools and techniques to support the tailoring process, for example: 

\faComment{}\textit{“The conventional agile practices must be used to develop the quantum software system. However, designing a compatible, customizable, and tailorable agile approach that defines the best practices for quantum software required a set of tools and technologies, and presently, the lack of such tools seems a major challenge”} [P8], Team lead.

\begin{itemize}
     \item \textbf{Continuous SE infrastructure}
 \end{itemize}
 Agile-driven quantum software development  infrastructure is required for implementing a deployment pipeline and monitoring dashboard to support the development experience \cite{khan2022embracing}, similar to modern-day continuous software engineering  \cite{fitzgerald2006customising}. This pipeline will collect data from the development activities and visualize the practitioner's experiences and customer feedback \cite{khan2022embracing}. However, such infrastructure is not yet implemented on a large scale, as mentioned by the interview participant:
 
\faComment{}\textit{“Agile-based quantum software development includes various phases which consist of multiple components and are often difficult to implement manually. It require infrastructure support complemented by open source tools to automate and enables continuous software engineering practices, e.g., continuous integration”} [P12], Software designer.

\subsubsection{Standards and Specifications}
Standards and specifications establish a common agreement for engineering criteria, terms, principles, items, practices, and processes \cite{Acq2021}. In this study, standardization and specifications emerged as the core category of challenging factors based on two elementary concepts: process standardization and optimum documentation.
\begin{itemize}
     \item \textbf{Process standardization}
 \end{itemize}
 Standardization plays an important role in portraying and strengthening the QC technologies \cite{artem2021standarization}. In line with it, the interview respondents perceived standardization as a roadmap for considering agile practices to develop a quantum software system. For instance, one of the interview participants described it as:
 
  \faComment{}\textit{"Using agile practices to develop a quantum software could be crucial to encapsulate the agile manifesto and develop a quality product. It is important to provide a roadmap (standards) that define rules, guidelines or characteristics for activities used to bring agile practices in quantum software development domain"} [P8], Team lead.
\begin{itemize}
     \item \textbf{Optimum documentation}
 \end{itemize}
 
 In agile software development,  optimum documentation provides the best possible process and the adaptability of changes across the  development life cycle \cite{Bendocumentation2016}\cite{rami2016documentation}. Simple and lightweight documentation in agile is “living,” and it needs to be collaboratively maintained by the whole team \cite{nuclino2016documentation}. Similar understanding (optimum documentation) is supported by the interview participants for agile-driven quantum software development. For instance, one of the interviewees mentioned: 

\faComment{}\textit{"Agile focus on minimizing waste; logically considering it, the project documentation is unnecessary and exhaustive activity. However, it doesn't mean documentation should be neglected and thrown away (particularly for hybrid agile-quantum activities). The new agile quantum software development genre must require documentation and guidelines at some extent to ensure the success of the process, product and project"} [P10], Scrum master.
\section{Discussion} \label{sec:discussion}
We now summarise the study's core findings (RQ1, RQ2), implications for research and practice and threats to the validity. 
\subsection{Agile for Quantum Software Development (RQ1)}
Applying agile practices enables QSE teams to tackle quantum software development endeavours by relying on iterative and incremental development to deliver quantum software. For instance, an increment of transforming the quantum domain knowledge (e.g., secure transmission of key in a quantum network) to a quantum algorithm (implementing Shor's algorithm for decryption) emphasizes a piecemeal development and quantum-specific knowledge being translated into implementation details \cite{zhao2020quantum}\cite{khan2022software}.

\begin{figure*}[!htbp]
 \centering
  \includegraphics[width=0.9\textwidth]{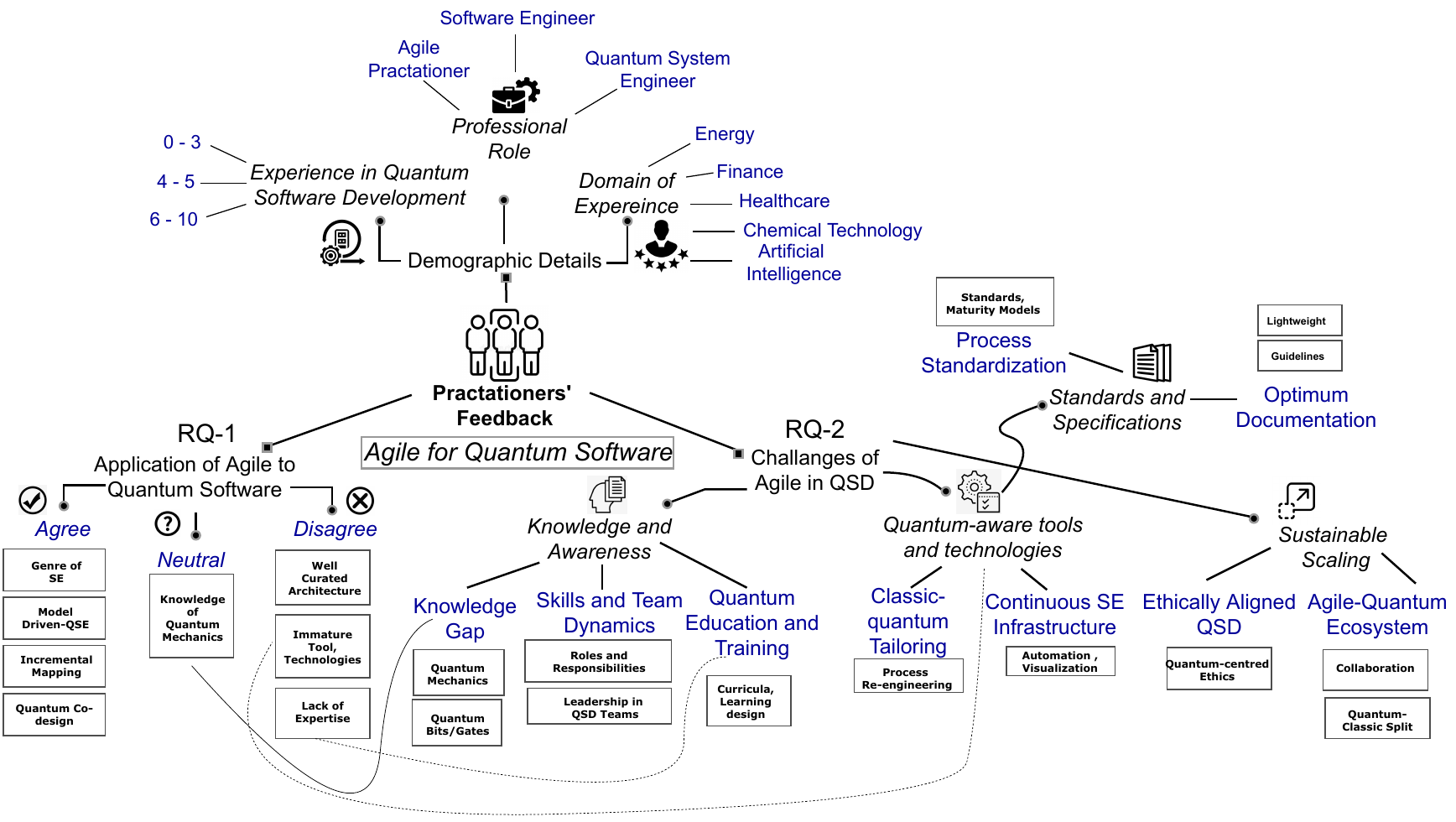}
 \caption{Discussion}
 \label{fig:discussion}
\end{figure*}

The overwhelming majority of the interviewed practitioners (i.e., 17/26) endorsed that the application of agile practices enables development teams to undertake quantum software development tasks more effectively and efficiently. Specifically, the practitioners rationalized their endorsement based on the view that QSE is yet another genre of SE based on quantum mechanics, therefore; the principle and practices of conventional agile practices can be applied in a quantum software development lifecycle \cite{hernandez2020quantum}\cite{piattini2021toward}\cite{khan2022embracing}. Moreover, the advocacy of agile for quantum software highlighted a multitude of reasons by practitioners such as model-driven QSE, increments to map various quantum-centric models (e.g., domain, architecture, code, simulation model), and quantum system co-design as shown in Figure \ref{fig:discussion}. On the contrary, approximately 1/4 (6/21) of the surveyed practitioners disagreed with agile-driven quantum software development sharing their skepticism as immature tools and technologies \cite{zhao2020quantum}, lack of professional expertise such as quantum software modeling \cite{gemeinhardt2021towards}, and the needs of a well-curated architecture \cite{ahmad2022towards}, among the reasons for the rejection (see Figure \ref{fig:discussion}). Three participants remained neutral, rationalising the reasons that the knowledge of quantum mechanics or lack of it may be determinantal factors for developing a team to opt for agile practices or not while undertaking quantum software projects (see Figure \ref{fig:discussion}). 

We noticed that the disagreement or neutral factors raised by the interview participants were later identified as the core categories of challenges. For example, the participants rejected using agile practices for developing quantum software because of immature tools and technologies, which is directly related to the \textit{Quantum-aware tools and technologies} category of challenges as shown in Figure \ref{fig:discussion}. It reveal that most of the disagreement and natural factors are  emerged as the core challenging categories (see Figure \ref{fig:discussion}).  

 
 


\subsection{Agile Practices Challenges in Quantum Software Development (RQ2)}
Agile SE to develop quantum software involves a multitude of challenges that vary from lack of knowledge and skills (people-centric), quantum-centric tools and development platforms (technology-centric), sustainable scaling (societal-centric ) along with models and documentation (process-centric) (see Figure \ref{fig:discussion}).

\underline{\textbf{Quantum-centric awareness and knowledge}}: The practitioners' responses reveal that lack of quantum-awareness in agile teams, i.e., limited or non-existent knowledge, skills, teamwork and education, are among the prime challenges for seamless and at-a-scale adoption of agile practices in the quantum software development context \cite{hernandez2020quantum}\cite{peterssen2020quantum}. The study analysis suggested that practitioners explicitly referred to knowledge as quantum mechanics know-how, such as quantum entanglement, quantum superposition etc. while manipulating quantum bits. In quantum software development, agile teams are perceived to work better when built-around practitioners who could be quantum-aware (team dynamics); this means that instead of a traditional project manager or scrum master a quantum domain engineer is perceived as the right professional to lead the software development tasks in a bottom-up approach from domain knowledge to quantum modelling, implementation, and execution/simulation \cite{ahmad2022towards}. Lack of education and knowledge are obvious challenges, however, it may be seen as an opportunity to design a course curriculum (degree level) that can help preparing the skilled workforce with extensive knowledge and understanding of agile-driven quantum software development. Such courses may help software engineers and developers to leverage their existing skill set to work on quantum software development \cite{peterssen2020quantum}\cite{li2021understanding}. 

\underline{\textbf{Ethically aligned agile-quantum ecosystem}}: In addition to the technical, socio-technical aspects faced by the quantum software team is perceived as a challenge with far-reaching consequences of what is referred to as the ignorance of ethics and the lack of agile-quantum ecosystem (section \ref{sec:sustainable scaling}). With the increased adoption of QC technologies, ongoing debates both in the academic and industry advocate for framework and guidelines for ethically-aligned quantum computing technologies \cite{saurabh2022ethical}. From an agile perspective, practitioners did emphasize that quantum-age software needs an ecosystem that provides services and processes for seamlessly bridging the gap between agile teams and stakeholders from other domains e.g., quantum physics and applied mathematics. Similarly, the computational supremacy of QC technologies in areas like security (e.g., data en-/decryption) and bio-inspired computing (e.g., gene editing) should adhere to development practices, audits, and metrics that ensure quantum software potential does not undermine social norms and values.

\underline{\textbf{Lack of maturity and openness in tools}}: Software tools such as model-to-code translators and test case generators are used to automate and customize developmental lifecycle tasks that may be time-consuming and error-prone \cite{zhao2020quantum}\cite{weder2021quantum}. One of such tasks is the conventional agile and Quantum tailoring \cite{hernandez2020quantum} that lacks the tool support, for example source code execution via classical (binary gates) and integration through quantum (Qgates). Another hindrance relates to the lack of infrastructure to visualize and automate the agile-driven continuous QSE activities. The availability of open-source tools can ensure community-wide initiatives to develop such infrastructure that can be customized, readily evolved, and widely adopted rather than individual solutions that may apply to a limited set of problems \cite{khan2022software}.

\underline{\textbf{Standards and optimal documentation}}: Agile based SE is seen as a light and adaptive mechanism to develop and deliver software-intensive systems and products \cite{beck2001manifesto}. In line with the agile manifestation for minimal documentation \cite{beck2001manifesto}, quantum software  standards can be treated as necessary instruments to define agile-aligned quantum software roles and essential documentation to guide the team. For example, a readiness model which provide a set of best practices and guidelines for classic-quantum transformation or hybridization (classic-quantum co-design) \cite{akbar2022classical}. In line with \cite{akbar2022classical}, standards  can act as  sufficient documentation and specifications that assist in moving from ad hoc agile-driven quantum software development practices to more mature processes.

 

 

\subsection{Implications for Research and Practice}\label{sec:implications}
Some implications arise from this research as listed in the following. 

\underline {\textbf{Implications for research}}: The results of the study complement the recently emerging streams of research on quantum software engineering, particularly add practitioner-centric view via empirically-based findings to agile-driven quantum software development. Researchers can utilize the results to quickly lookup the streamlined challenges (i.e., new hypotheses to be tested) regarding the knowledge, skills, tools, technologies, ethics, sustainability, and process aspects of quantum software engineering. The study’s findings, on the other hand, give quick access to the body of knowledge based on the practitioner’s understanding of agile-driven quantum software development.

 \underline{\textbf{Implications for practice}}: The study delivers inferential empirical findings to industrial practitioners and/or stakeholders interested in the fine-grained analysis of agile practices for quantum software development. The overview and interpretations of adaptability and potential challenges of agile-driven quantum software development provided a roadmap enlightening software practitioners interested in using agile practices for developing quantum software. In particular, the practitioners perspectives could help developers to engineer the next generation of quantum software.

\subsection{Threats to validity} \label{sec:threats}
Various potential threats could affect the validity of the study findings. The relevant threats are broadly categorized across internal, construct, external and conclusion validity \cite{zhou2016map}. 

\underline{\textbf{Internal validity}}: Internal validity is the extent to which particular factors affect the methodological rigor. In this study, the first threat to internal validity is the interview participant’s understanding of the interview questions. This threat has been mitigated by conducting pilot interviews to ensure the understandability and readability of the interview questions (see section \ref{sec:InterviewMetho}). The limited or no relevant expertise of participants to answer the interview questions is also a potential internal validity threat. We approached agile SE and QC practitioners using the personal network, industrial collaboration,  code hosting platforms (e.g., GitHub) and professional social media networks (e.g. LinkedIn). Only participants having mutually inclusive knowledge of conventional agile  and QC practices were selected. The interpersonal bias in the data collection process may threaten the internal validity of study findings. We mitigated this threat by organizing regular consent meetings between all the authors for interview instrument development, feedback and GT data analysis (i.e. coding, concept development, categorization). GT findings are limited and applicable to the studied contexts and domains  \cite{glaser2017discovery},which in succession are dictated by access to the study participants. Therefore, we strongly ensured that all the details of interview participants- individual to organizational- will be strictly confidential. The study findings would be enormously used for research purposes only and not be shared with third-party under any circumstances.

\underline{\textbf{Construct validity}}: Construct validity is the extent to which the study constructs are well defended and interpreted. In this study, the perceptions of the interview participants on agile adaptability and relevant challenges are the core constructs. The verifiability of constructs is a known limitation of GT studies. It could be inferred from the research method efficacy and from the evidence that the study findings are presented based on the data collected using the selected research method \cite{hoda2012developing}. Therefore, we explicitly discussed the step-by-step process of the research method (see section \ref{sec:Analysing Interview Data}), the defined categories supported with quotes from the interview participants and our observations in section \ref{sec:ResultsRQ2}. It exhibits how the systematically defined research protocol  and reported findings (RQ1, RQ2) support the verifiability of  the study constructs.

\underline{\textbf{External validity}}: External validity refers to broadly generalizing the study findings in other contexts. In this study, the sample size and sampling approach may not provide a strong foundation to generalize the findings. However, it is know fact that QSE, and particularly using agile practices for developing quantum software, are new research areas and not in practice at a high level. We tried to mitigate this threat by using all possible sources (see section \ref{sec:InterviewMetho}) to approach the potential population. We collected data from 10 countries across 4 continents, with participants having 16 different professional roles and a diverse range of experience in various industrial domains (see Figure \ref{fig:demography}). Moreover, we plan to extend this study in the future with a large data sample from multiple data sources (i.e. mining the Q\&A platforms, conducting industrial surveys and interviews) (see section \ref{sec:conclusions}). 

\underline{\textbf{Conclusion validity}}: Conclusion validity refers to the factors that affect the credibility of the study conclusions. To address this threat, the first three authors were continuously involved in the interview instrument development and data analysis process. The second author mainly collected the data by conducting live Zoom sessions with the participants. However, all other authors occasionally (consent meetings) reviewed the data and provided their feedback to tackle the conflicts that appeared during the data collection and analysis process. Finally, brainstorming sessions were conducted, and all the authors participated in discussing the study findings and drawing conclusions.

\section{Related Work}
\label{sec:relatedWork}
We discuss the most relevant existing work, reviewing state-of-the-art on processes and development lifecycle(s), for quantum software systems \cite{zhao2020quantum}. A conclusive summary at the end reiterates the scope and contributions of the proposed research. 

\subsection{Process-centred Engineering of Quantum Software}
Quantum software engineering – most recent genre of software engineering – aims to apply existing processes, practices, methods, tools, and techniques to design and develop quantum age software systems and applications effectively and efficiently \cite{ali2022software}\cite{piattini2021toward}. The existing body of SE knowledge provides foundations for quantum software development \cite{zhao2020quantum}, however; unique characteristics of quantum software (e.g., quantum system co-design, quantum source coding) and quantum specific professional expertise (e.g., quantum domain engineer, quantum source coder) requires traditional SE methods to be tailored to address QSE challenges \cite{ahmad2022towards}\cite{khan2022software}. To investigate quantum specific software architecting and development, recently conducted systematic reviews aim to establish empirical foundations to design \cite{khan2022software} and develop \cite{heim2020quantum} quantum software by streamlining the required process, existing patterns, and ideal tool chain(s) that can empower the roles of traditional software developers as quantum software architect and developers. QSE as a discipline is still in its infancy, however; community wide initiatives are gaining momentum, for example with dedicated conference\footnote{\url{https://conferences.computer.org/qsw/2022/}}, workshop (Q-SE)\footnote{\url{https://conf.researchr.org/home/q-se-2022}}, literature reviews \cite{heim2020quantum}\cite{zhao2020quantum}\cite{gill2022quantum}\cite{khan2022software}, and repositories mining \cite{de2022software} efforts with a common denominator of using best practices and reusable knowledge (processes, patterns, reference architectures, etc.) to effectively and efficiently develop quantum software \cite{ali2022software} 

\subsection{Iterative Development of Quantum Software}
To synergise an iterative development with quantum software engineering process, Khan et al. \cite{khan2022embracing} present the vision for continuous iterative development and delivery of quantum software. However, iterative enabled quantum software development is still in its infancy and demands novel domain specific techniques and processes as suggested in \cite{khan2022embracing}\cite{hernandez2020quantum}\cite{weder2021quantum}. Considering the existing QSE methods and techniques\cite{weder2021quantum}\cite{gemeinhardt2021towards}\cite{wang2022generating}\cite{ali2020modeling} there is still no or even consensus on agility in quantum software. The adoption of agile for quantum software development can help practitioners to iteratively experiment new ideas, and improve performance- all in parallel \cite{khan2022embracing}\cite{hernandez2020quantum}. In line with this, Gonzalez and Paradela \cite{hernandez2020quantum} conceptualised a hybrid project management framework (unifying classical and quantum SE) for agile and incremental management of software that can be adapted to the needs of quantum-age computing. However, the study do not presented any empirical evidence to actualise the proposed framework. Similarly, using the contemporary SE practices, Weder et al. \cite{weder2021quantum} introduces a quantum-based software development process that comprises of eight classical phases with induction of a new phase called quantum-classical splitting. The quantum-classical splitting phase is tailored for quantum software systems and it distinguishes between parts of the quantum software that need to be executed on a quantum computer and that ones that can be executed on classical computers. 

\subsection{Conclusive Summary}

In this research, we argue that prior to adopting any specific processes and/or reference models \cite{weder2021quantum} \cite{hernandez2020quantum} for agile-driven QSE, there is a need to analyse practitioners perceptions, professional practices, and empirically grounded findings on the potential and challenges of adopting agile SE in the context of quantum software. Our research aims to complement the existing efforts, such as establishing models \cite{gemeinhardt2021towards}, architectures \cite{khan2022software}, and processes \cite{moguel2020roadmap} with an overall aim to systemize the development of quantum software. This research follows-up on our previous works (i.e., the cases of architecting \cite{ahmad2022towards}\cite{khan2022software} and iterating \cite{khan2022embracing} quantum software development) with a practitioner’s interview on the potential of adopting agile practices for quantum software development.
\section{Conclusions and Future Work}
\label{sec:conclusions}



Quantum software engineering leverages the principle and practices of classical SE- empowering the role of architects and developers to deliver quantum age software applications- providing impetus to QC. We conducted this study, rooted in empirical analysis and grounded theory, to understand practitioners’ perspectives on the potentials and pitfalls of agile practices in quantum software development endeavours. We engaged a total of 26 practitioners from 10 countries generally classified as software engineers, agile experts, and quantum software developers ( see Figure \ref{fig:demography}) with an overwhelming majority agreeing (n=17), approximately one fourth disagreeing (n=6) and relative minority (n=3) not sure on applying agile practices to quantum software development, answering RQ1. 

To gain better understanding and fine-granular analysis of the disagreements, we presented open-ended questions on challenges, i.e., hindrances perceived by practitioners on agile-oriented quantum software development, answering RQ2. After collecting and synthesizing the responses, we identified 4 categories organized into 9 sub-categories (concepts) perceived as challenges by practitioners (see Figure \ref{fig:discussion}). These four categories include: \textit{1) Knowledge and awareness}, which provide practitioners insights and understanding of challenges that could hinder the adaptability of agile practices for developing quantum software. For instance, the existing knowledge gap between classical and quantum software development, different mindsets (classical->quantum), and lack of agile-quantum specific education and guidelines.\textit{2) Sustainable scaling}, encapsulates the challenges related to harming software sustainability practices. QSE is a paradigm shift, and using agile to develop quantum software is more likely to threaten the existing ethical protections (e.g., transparency, accountability) and the agile-quantum ecosystem. \textit{3)  Quantum-aware tools and technologies}, category consist of the challenges related to the immature tool support and lack of infrastructure to tailor, customize, automate, and configure the agile practices for developing quantum software. \textit{4) Standards and specifications}, category highlights the lack of common rules and principles for agile-quantum software scenario. From the perspective of participants, to adopt agile for quantum software the necessary standards, models, and documentation should be provided.

Our future work is to extend this study by identifying the challenges from social coding platforms in open-source quantum projects. To achieve this, we plan to 1) mine publicly available developers’ discussions and code repositories (i.e., Stack Overflow, GitHub) to explore solutions that the QC community employs for the effective adoption of agile methods, 2) map these solutions to the identified challenges with respect to causes, and 3) validate this mapping via a large sample quantitative survey and interviews of QSE practitioners.

\section*{Acknowledgments} \label{sec:ack}
The research team would like to thank the interview participants and industrial partner who generously participated and helped us in this research and shared their time and experience. 
\balance

\bibliographystyle{ACM-Reference-Format}
\bibliography{main}

\end{document}